\documentclass[conference]{IEEEtran}
\IEEEoverridecommandlockouts
\usepackage{cite}
\usepackage{amsmath,amssymb,amsfonts}
\usepackage{algorithmic}
\usepackage{graphicx}
\usepackage{textcomp}
\usepackage{xcolor}

\usepackage{caption}
\usepackage{subcaption}
\usepackage{multirow}

\def\BibTeX{{\rm B\kern-.05em{\sc i\kern-.025em b}\kern-.08em
		T\kern-.1667em\lower.7ex\hbox{E}\kern-.125emX}}
\begin{document}
	
	\title{TauPETGen: Text-Conditional Tau PET Image Synthesis Based on Latent Diffusion Models  
	}
	
	\author{Se-In Jang,  Cristina Lois, Emma Thibault,  J.  Alex Becker, Yafei Dong, \\ Marc D. Normandin, Julie C. Price, Keith A. Johnson, Georges El Fakhri, and Kuang Gong\textsuperscript{*} 
	\thanks{The authors are with the Department of Radiology, Massachusetts General Hospital and Harvard Medical School, Boston, MA, 02114, USA (email\textsuperscript{*}:kgong@mgh.harvard.edu).}
	}

	\maketitle

\begin{abstract}
In this work, we developed a novel text-guided image synthesis technique which could generate realistic tau PET images from textual descriptions and the subject's MR  image.  The generated tau PET images have the potential to be used in examining relations between different measures and also increasing the public availability of tau PET datasets.  The method was based on latent diffusion models.  Both textual descriptions and the subject's MR prior image were utilized as conditions during image generation.  The subject's MR image can provide anatomical  details,  while the text descriptions,  such as gender,  scan time,  cognitive test scores, and amyloid status,  can provide further guidance regarding where the tau neurofibrillary tangles might be deposited.  Preliminary experimental results based on clinical [18F]MK-6240 datasets demonstrate the feasibility of the proposed  method in generating realistic tau PET images at different clinical stages.
\end{abstract}


\section{Introduction}
Tau PET imaging is an essential technique for the  diagnosis and treatment monitoring of neurodegenerative diseases. It enables {\it{in vivo}} visualization of tau neurofibrillary tangles deposits in the brain.  Currently accessibility of tau PET imaging is  limited to research studies and clinical trials,  and the availability of public tau PET datasets is limited.  Recently,  image synthesis techniques have achieved great success in generating realistic images that can be used in different areas, such as treatment planning and PET attenuation correction.  In this work,  we aimed to increase the availability of tau PET datasets and also better examine relations among different cognitive-test and imaging measures through image synthesis. 

Several approaches were proposed for image synthesis in the context of medical imaging, such as generative adversarial networks (GAN) and variational autoencoders (VAE).  GAN was widely used for medical image synthesis but often suffered from mode collapse and training instability.  VAE, on the other hand, could generate diverse images but struggled to maintain the high quality of generated images. 
Diffusion Models (DM) \cite{sohl2015deep, ho2020denoising} were a class of generative models that learned to model the data distribution by simulating a diffusion process. The core idea was to train a denoising function to reverse a diffusion process from a noisy state to the original data point. 
Latent diffusion models (LDM) \cite{rombach2022high} recently emerged as a promising alternative for high quality image generation with a more controllable synthesis process (i.e., with text).
LDM had been successfully applied to natural image synthesis, but their application to medical imaging, particularly PET imaging, remained largely unexplored. 

In this work, we presented \textit{TauPETGen}, a novel text-conditional image synthesis approach  based on LDM, which could generate realistic tau PET images based on textual descriptions. Our approach aimed to provide researchers and clinicians with a tool to better understand disease progression and also increase the availability of public tau PET datasets.
\begin{figure*}[!t]
	\centerline{\includegraphics[width=2\columnwidth]{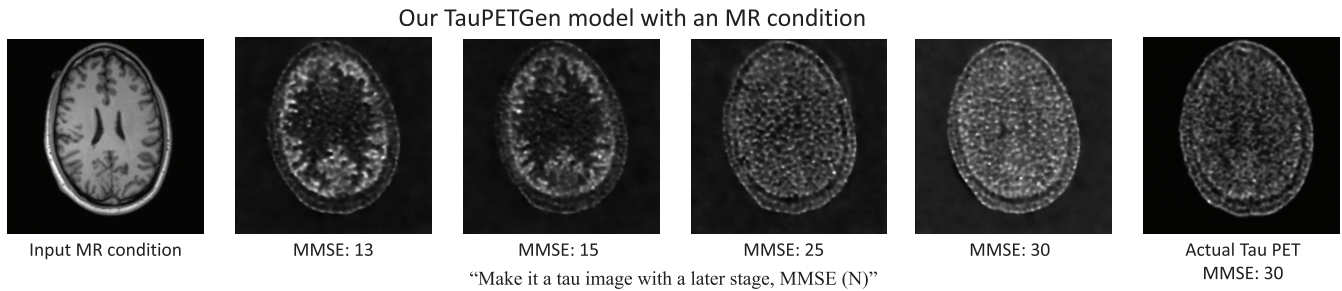}}
	\caption{\small{The generated synthetic tau PET images with  the subject's MR image (left column) with varying MMSE values as the text input.  Format of the text input was shown in the caption.  Here later stage represented 90-110 min post injection.}} 
	\label{fig.exp1}
\end{figure*}
\begin{figure}[!t]
	\centerline{\includegraphics[width=\columnwidth]{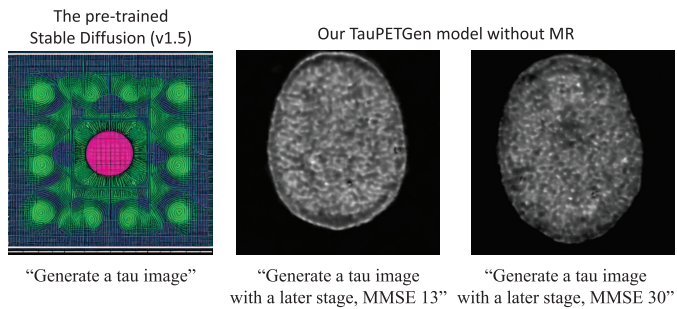}}
	\caption{\small{Synthetic tau PET images generated based on the Stable Diffusion model and the TauPETGen model without MR as the input.}}
	\label{fig.exp2}
\end{figure}

\section{Proposed Method}
The core idea of the proposed method was to model the image generation process as a controlled diffusion process in a latent space. Textual descriptions were encoded into this latent space, and the encoded information guided the diffusion process to synthesize the desired tau PET image. The LDM's loss function, when accompanied by a text-condition $c_{\text{text}}$,  could be expressed as:
\begin{equation}
	\mathcal{L}(\theta) =  \mathbb{E}_{\mathcal{E}(x), \epsilon \sim \mathcal{N}(0, 1), t} 
	\left[|| \epsilon - \epsilon_{\theta}\left(z_t, t, c_{\text{text}}\right)  ||^2_2 \right],
\end{equation}
where  $\theta$ represented the parameters of the denoising function, and $x$ was the tau PET image.
The encoder $\mathcal{E}(\cdot)$ transformed $x$ into a latent representation $z_t=\mathcal{E}(x)$, and the decoder $\mathcal{D}(\cdot)$ reconstructed the tau PET image from the latent. $z_t$ was an encoded noisy latent of the tau PET image, while the noisy level increased over the time steps $t \in T$.  
$\epsilon_{\theta}(\cdot)$ was the denoising autoencoder as a time-conditional Unet.
The final genereated tau image was $\tilde{x} = \mathcal{D}(\mathcal{E}(x))$.

The LDM presented a promising approach for synthetic image generation. Nonetheless, anatomical consistency in the final generated tau PET image was not guaranteed. Consequently, the text-conditional model could potentially generate tau PET images with  inaccurate anatomical structures if the number of training datasets was not large enough, which was always the case for tau PET imaging. 
Simialr to the instructPix2Pix model \cite{brooks2022instructpix2pix}, we trained the network, denoted as $\epsilon_{\theta}(\cdot)$,  by adopting a loss function that incorporated the subject's MR image $c_{MR}$ as an additional condition, which helped ensure anatomical consistency:
\begin{equation}
	\mathcal{L}(\theta) =  \mathbb{E}_{\mathcal{E}(x), \mathcal{E}(c_{\text{MR}}), \epsilon \sim \mathcal{N}(0, 1), t} 
	\left[|| \epsilon - \epsilon_{\theta}\left(z_t, t, \mathcal{E}(c_{\text{MR}}), c_{\text{text}}\right)  ||^2_2 \right].
\end{equation}
The concatenation of $z_t$ and $\mathcal{E}(c_{\text{MR}})$ as the input to the first convolutional layer allowed the network to use both information, and could potentially help the network preserve the anatomical structure present in the MR prior image.

\section{Experiments}

\subsection{Settings}

A total of 139 pairs of [18F]MK-6240 PET and MR images were utilized, each accompanied by its corresponding textual descriptions. The tau PET scans were acquired from GE DMI PET/CT scanner. The tau images reconstructed from 90-110 mins post injection were denoted as \textit{the later stage} tau PET images, and those reconstructed from 0-10 mins were denoted as \textit{the early stage} tau PET images.  The Mini Mental State Examination (MMSE) score was utilized as the cognitive test measure.  Since each subject may have different MMSE values, each tau PET image was accompanied by a specific description that included the corresponding MMSE value. For example, the descriptions included phrases like ``\textit{a tau image with later stage, MMSE 30}'' and ``\textit{a tau image with later stage, MMSE 13.}'' This allowed the text-conditional network model to correlate the cognitive status of the subject, as indicated by the MMSE score, with the corresponding tau PET image. We employed the pre-trained Stable Diffusion (v1.5) model \cite{rombach2022high} as a reference for comparison. The Stable Diffusion \cite{rombach2022high} and InstructPix2Pix \cite{brooks2022instructpix2pix} architectures were kept consistent for TauPETGen, both in the absence and presence of the MR condtiion. Our TauPETGen models were trained with 10k steps based on 2D images.

\subsection{Results}
Fig. \ref{fig.exp1} shows the synthetic tau PET images generated by the proposed \textit{TauPETGen} model with the subject's MR image and varying MMSE scores.  For the true tau PET image shown in the right column, there was no obvious cortical tau uptakes as the subject was healthy.   As the text condition changed, i.e., MMSE values decreased, the generated tau PET images had increased uptakes in the cortical regions.  Fig. \ref{fig.exp2} presented an ablation study comparing the pre-trained Stable Diffusion and TauPETGen using only text condition. When prompted with ``\textit{generate a tau image,}'' the Stable Diffusion produced an entirely incorrect tau image, possibly due to a lack of knowledge about tau PET imaging. Our TauPETGen created tau PET images with the prompt ``\textit{generate a tau image with a later stage, MMSE (13 and 30),}'' but the images appeared inaccurate and lacked an understanding of the anatomical structure.  The proposed method outperformed the text-only conditioned models, i.e., Stable Diffusion and TauPETGen without MR, by preserving anatomical structures while also yielding more accurate uptake patterns at different clinical stages.

\section{Conclusion}
In this work, we proposed a text-conditional image synthesis approach for tau PET imaging based on latent diffusion models.  The proposed method generated realistic images based on textual descriptions and had better preservation of anatomical structures with the additional MR condition. Experimental results demonstrated the feasibility of our approach in generating high-quality tau PET images, highlighting its potential for synthetic tau PET image generation with different clinical measures as a reference check,  and enlarging the public available tau data pool.  Our future work will focus on further improving the synthesis quality by incorporating additional prior images and clinical measures.

\bibliographystyle{IEEEtran} 
\bibliography{references.bib}

\end{document}